\documentclass[preprintnumbers,amsmath,amssymb,nofootinbib]{revtex4}

\usepackage{graphicx}
\usepackage{epsfig}
\usepackage{color}

\newcommand{\beq}{\begin{equation}}
\newcommand{\eeq}{\end{equation}}
\newcommand{\bea}{\begin{eqnarray}}
\newcommand{\eea}{\end{eqnarray}}
\newcommand{\bwd}{\begin{widetext}}
\newcommand{\ewd}{\end{widetext}}

\begin{document}

\title{Fast longitudinal beam dynamics optimization in x-ray FEL linear accelerators}


\author{Ji Qiang}
\email{jqiang@lbl.gov}
\affiliation{Lawrence Berkeley National Laboratory, Berkeley, CA 94720, USA}

\begin{abstract}
A high peak current, flat longitudinal phase space electron beam is desirable
for efficient x-ray free electron laser (FEL) radiation in next
generation light sources. To attain such a beam requires the 
extensive design of the linear accelerator (linac)
including both linear and nonlinear effects. In this paper,
we propose a lumped longitudinal beam dynamics model for fast optimization of
the electron beam longitudinal phase space through the accelerator. 
This model is much faster than available tracking programs and also
shows good agreement with the fully three-dimensional element-by-element multi-particle simulations.
We applied this model in a parallel multi-objective differential evolution optimization program
to an existing LCLS-II superconducting linac design and obtained an optimal solution with
significantly higher core peak current than the original design. 
\end{abstract}

\maketitle

\section{Introduction}

High brightness, coherent x-ray radiation from a free electron laser (FEL) light source
provides an important tool for scientific discovery in physics, chemistry,
biology and other fields. 
To produce such a coherent radiation effectively requires the use of 
high brightness electron beam with high peak current, small energy spread, and 
small emittance
inside the radiation undulator.
The high brightness electron beam used in an x-ray FEL light source typically comes from
a linear accelerator beam delivery system~\cite{lcls,fermi,xfel,jfel,palfel,swissfel}.
This system consists of a photoinjector to generate an initial high brightness electron beam,
a RF linac to accelerate the beam to multiple GeV energy and to compress the beam
to hundred or thousand Ampere peak current, and transport beam line to deliver
the beam to different undulators.

To attain a high brightness electron beam at the entrance of undulator involves
extensive design and optimization of the linear accelerator parameters.
In previous studies, analytical model was developed to choose the settings
of linac accelerating gradients, phases, and magnetic chicane parameters~\cite{zagorodnov,dimitri}. 
This model gives a quick estimate of the RF cavity accelerating gradient amplitude, phase
and bunch compressor parameter settings inside the linac but it
does not include collective effects such as space-charge effects,
coherent synchrotron radiation (CSR), and 
structure or resistive wall wakefields inside the accelerator.
It works well when the peak current is not very high and the collective
effects are weak. It can be used as a starting point of longitudinal phase
space optimization. However, when one pushes the limit of final core peak
current, those collective effects are no longer negligible and can have
major impact on electron beam longitudinal phase space distribution.
This is especially true after the final stage of beam compression, the
electron bunch length becomes as short as a few tens microns. Using the phase
setting in the RF cavities to remove the correlated energy spread (also called chirp) of the 
electron beam becomes inefficient. Meanwhile, given the high peak current
after compression, the collective effects such as wakefields 
can become dominant.
For example, in the LCLS-II design, the final correlated energy spread
of the electron beam is
almost completely removed by the resistive wall wakefields before
entering the undulator~\cite{paul,tor}. Including those collective effects 
in the longitudinal
beam dynamics model is crucial for the high peak current accelerator design.

To include those effects, one normally resorts to detailed multi-particle element-by-element tracking
simulations.
In the past studies, a number of tracking codes such as LiTrack~\cite{litrack}, 
Elegant~\cite{elegant}, and
IMPACT~\cite{impact1,impact2} were used for element-by-element electron linac simulations.
Among those three codes, LiTrack handles only longitudinal beam dynamics and
has the fastest computational speed. However, LiTrack does not include the CSR effect and
the longitudinal space-charge effect in simulation.
The other two codes include all those collective effects, but 
are much slower in computational speed due to the use of fully three-dimensional element-by-element multi-particle tracking.
In this paper, we propose a lumped longitudinal beam 
dynamics model for fast longitudinal phase space optimization.
This model is based on the assumption that out of the injector, inside the RF linac, when the
electron beam energy is sufficiently high (e.g. $\ge 100$ MeV),
the transverse beam dynamics and the longitudinal beam dynamics can be 
reasonably decoupled.
The transverse focusing elements such as quadrupoles are treated
as drifts in the longitudinal beam dynamics model. The RF cavities 
in each section of the linac are lumped
as one RF accelerating element. The magnetic bunch compression chicane is treated
as a thin lens element except that the last dipole bending magnet is used
for the CSR effect calculation. The longitudinal space-charge effect, the
structure wakefields from the RF cavities, and the resistive wall wakefields
from the transport beam line pipe are included in the model. 
Using the lumped elements
significantly improves the speed of simulation while including all longitudinal
collective effects. In the simulation, those collective effects are
computed efficiently using an FFT based method.
Using a weighted macropaticle method, given the initial current profile
and correlated energy profile in longitudinal phase space, 
a small number of macroparticles 
(from a few hundreds to a thousand macroparticles)
that corresponds to the longitudinal slice coordinates are needed
in the simulation. All these make the longitudinal beam dynamics simulation very
fast.

The organization of this paper is as follows: after the Introduction, we present the
one-dimensional (1D) longitudinal beam dynamics model in Section II;
We discuss about a recently developed multi-objective differential evolution optimization
algorithm in Section III; We apply the multi-objective longitudinal beam dynamics optimization to an existing
LCLS-II design in Section IV;
and draw conclusions in Section V.

\section{Longitudinal beam dynamics model}

In this study, we focus only on the longitudinal beam dynamics,
and neglect the transverse motion of electrons.
Each electron macroparticle has longitudinal coordinates $(z,\Delta \gamma)$
with respect to the reference particle $(s_0,\gamma_0)$
and charge weight $w$. Here, $z =s-s_0$ is the bunch length coordinate ($z_{max}$
corresponds to the bunch head and $z_{min}$ the bunch tail), 
$\Delta \gamma  = \frac{E - E_0}{mc^2}$,
$E$ is the total energy of the particle, $E_0$ is the total energy of the
reference particle, $m$ is the rest mass of the particle, and $c$ is the speed of light in vacuum. 
For the longitudinal beam dynamics study, we include only drifts, RF cavities, and magnetic
compression chicanes as the beam line elements of the x-ray FEL linear accelerator. 
The other focusing elements such as quadrupoles are treated as drifts too.

For a macroparticle transporting through the lumped RF cavity element with total length $L_{acc}$, 
its longitudinal coordinates will be updated by the following equations:
\begin{eqnarray}
z & = & z + \frac{L_{acc}}{2}\Delta \gamma /(\gamma_0 \beta_0)^3   \\
\gamma_0 & = & \gamma_0 + \frac{L_{acc}}{2} V_{acc} \cos(\phi_0)  \\
\Delta \gamma & = & \Delta \gamma + L_{acc} V_{acc} (\cos(\phi_0-k z) - \cos(\phi_0))  \\
z & = & z + \frac{L_{acc}}{2}\Delta \gamma /(\gamma_0 \beta_0)^3   \\
\gamma_0 & = & \gamma_0 + \frac{L_{acc}}{2} V_{acc} \cos(\phi_0)  
\end{eqnarray}
where $V_{acc} = q V_{rf}/L_{acc}$ is the accelerating gradient amplitude, $k$ is the RF wave number,
and $\phi_0$ is the RF cavity design phase.

The magnetic bunch compression chicane is modeled as a thin lens element.
The particle longitudinal position through the chicane is given by~\cite{r56}:
\begin{eqnarray}
	z & = & z + R_{56} \frac{\Delta \gamma}{\gamma_0} + T_{566} (\frac{\Delta \gamma}{\gamma_0})^2
	+ U_{5666} (\frac{\Delta \gamma}{\gamma_0})^3
\end{eqnarray}
where
\begin{eqnarray}
	R_{56} & \approx &  2 \theta^2(L_{db} + \frac{2}{3} L_b) \\
	T_{566} & \approx & -\frac{3}{2} R_{56} \\
	U_{5666} & \approx & 2 R_{56}
\end{eqnarray}
where $\theta$ is the bending angle of one of dipole magnets (assuming that
all four dipoles have the same bending angle amplitude), $L_b$ is the 
length of the dipole magnet, and $L_{db}$ is the 
distance between
the first and the second (or between the third and fourth) dipole bending magnets.
From our benchmark with fully 3D element-by-element tracking using $5^{th}$ order
transfer map for the dipole bending magnet, we need to increase the $R_{56}$ 
by $0.5\%$ in order to match the current profile after the chicane with that from the 3D model.

Collective effects such as longitudinal space-charge effect, structure and resistive wall wakefields, and coherent 
synchrotron radiation play an important role in the longitudinal beam dynamics and are included
in this model. For the longitudinal space-charge effect, instead of using the space-charge impedance
model in the frequency domain~\cite{impact2}, we assume that the electron beam is
a round cylinder with separable uniform transverse density distribution and longitudinal density distribution. 
The longitudinal space-charge field on the axis is given as:
\begin{eqnarray}
	E_z^{sc}(0,0,z) & = & \frac{1}{4\pi \epsilon_0}\frac{2}{a^2}\int \frac{\gamma_0(z-z')\rho(z')}{(\gamma_0^2(z-z')^2+r'^2)^{3/2}} r' dz'dr' 
\end{eqnarray}
After integrating with respect to the transverse radial dimension, the longitudinal
space-charge field on the axis can be written as:
\begin{eqnarray}
	E_z^{sc}(z) & = & \frac{1}{4\pi \epsilon_0}\frac{2}{a^2}{\Big (}\int_{z_{min}}^z \rho(z') dz' - \int_z^{z_{max}}  \rho(z') dz' - \int_{z_{min}}^{z_{max}} \frac{\gamma_0(z-z')\rho(z')}{\sqrt{\gamma_0^2(z-z')^2+a^2}} dz'{\Big )}
\end{eqnarray}
where $a$ is the radius of the cylinder, $z_{min}$ and $z_{max}$ denote the minimum and the maximum
longitudinal bunch length positions, and $\rho$ is the electron beam longitudinal charge density distriubtion.
The above convolution can be computed efficiently using an FFT based method~\cite{hockney,impact3}.

The longitudinal wakefields from both the structure wakefields of RF cavities and the resistive wall
wakefields are included in the model. The longitudinal field from the wakefields are calculated
from the following convolution:
\begin{equation}
	E_z^{wk}(z) = \int_z^{z_{max}} W_L(z-z') \rho(z') d z'
\end{equation}
where $W_L(s)$ is the longitudinal wake function.
This convolution can also be computed efficiently using the FFT based method~\cite{impact2}.

The coherent synchrotron radiation through a bending magnet can be calculated from
the following integral:
\begin{equation}
	E_z^{csr}(z) = \int_{z_{min}}^z W_{csr}(z,z') \rho(z') d z'
\end{equation}
where $W_{csr}(s)$ is the longitudinal CSR wake function 
that includes both transient and steady-state radiations
through a bending magnet following Saldin et al. (Case A-D)~\cite{csr0}.
The CSR wake function has a very sharp variation around the origin
of bunch length coordinate. To avoid the use of large number of numerical
grid points to resolve the sharp variation,
instead of using the original CSR wake function and calculating the above integral directly, 
we divide this integral into the summation of a number of small interval integrals
and rewrite the integral for the CSR wakefield at location $z_k$ as:
\begin{equation}
	E_z^{csr}(z_k) = \sum_{k'=1} \int_{z_{k'}-h/2}^{z_{k'}+h/2} W_{csr}(z_k,z') \rho(z') dz'   
\end{equation}
If we assume that the longitudinal density $\rho(z')$ is constant within that small
interval, 
the above integral at $k$ slice can be approximated as~\cite{csr1,csr2}:
\begin{equation}
	E_z^{csr}(z_k) = \sum_{k'=1} \rho(z_{k'}) w_{k,k'}   
\end{equation}
where
\begin{eqnarray}
	w_{k,k'} & = & \int_{z_{k'}-h/2}^{z_{k'}+h/2} W_{csr}(z_k,z') d z' \\
	& = & I_{csr}(z_k,z_k'-h/2) - I_{csr}(z_k,z_k'+h/2)
\end{eqnarray}
where 
\begin{equation}
	I_{csr}(z,z_t) = -\int_{z_{min}}^{z_t} W_{csr}(z,z')dz'
\end{equation}

In case A (transient at entrance) where radiation source electron is 
in front of bending dipole magnet while the observer electron is inside
the magnet, the integrated CSR wake function is given as:
\begin{eqnarray}
I_{csr} & = &\frac{\gamma r_c mc^2}{R}\Big(\frac{1}{\zeta}-\frac{2(\hat{\phi}+\hat{y})+\hat{\phi}^3}{(\hat{\phi}+\hat{y})^2+\hat{\phi}^4/4} \Big ) 
\end{eqnarray}
where $\hat{\phi}=\gamma \phi$, $\phi$ is the azimuthal angle of the observer electron,
$\zeta = (z-z_t)\gamma^3/R$, $R$ is the bending radius of the dipole,
and $\hat{y}$, the normalized distance from the source to the dipole entrance,
can be found from the solution of the following equation:
\begin{equation}
\zeta = \frac{\hat{\phi}+\hat{y}}{2}+ \frac{\hat{\phi}^3}{24}\frac{\hat{\phi}+4\hat{y}}{\hat{\phi}+\hat{y}}
\end{equation}

\begin{figure}[!htb]
   \centering
   \includegraphics*[angle=0,width=300pt]{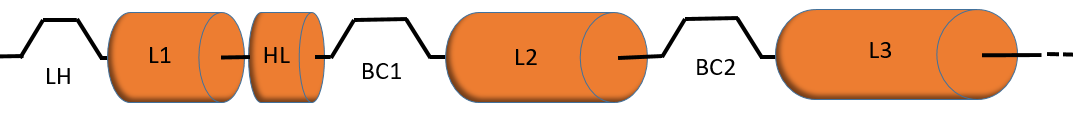}
	\caption{A schematic of LCLS-II linac.}
   \label{fig0}
\end{figure}

In case B (steady-state) where both the source and the observer are inside
the dipole, the integrated CSR wake function is given as:
\begin{eqnarray}
I_{csr}
	& = &-\frac{\gamma r_c mc^2}{R}\Big(\frac{4(\hat{u}^2+8)\hat{u}}{(\hat{u}^2+4)(\hat{u}^2+12)} \Big )
\end{eqnarray}
where $\hat{u}$, the normalized angle between the observer and the source,
can be found from the solution of the following equation:
\begin{equation}
\zeta = \frac{\hat{u}^3}{24}+ \frac{\hat{u}}{2} 
\end{equation}

In case C (transient at exit) where the source is in front of the dipole
entrance while the observer is outside the dipole exit, the integrated CSR wake function is
given as:
\begin{eqnarray}
I_{csr} & = &\frac{\gamma r_c mc^2}{R}\Big( \frac{1}{\zeta}-\frac{2(\hat{\phi}_m+\hat{x}+\hat{y}+\hat{\phi}_m^3/2+\hat{\phi}_m^2\hat{x})
	}{(\hat{\phi}_m+\hat{x}+\hat{y})^2+(\hat{\phi}\hat{x}+\hat{\phi}_m^2/2)^2} \Big )
\end{eqnarray}
where $\hat{\phi}_m = \gamma \phi_m$, $\phi_m$ is the total bending angle of the dipole, 
$\hat{x} = \gamma x/R$, $x$ the distance from the dipole exit to the observer, and
the normalized distance from the source to the dipole entrance
$\hat{y}$ can be found from the solution of the following equation:
\begin{equation}
	\zeta = \frac{\hat{\phi}_m+\hat{x}+\hat{y}}{2}+ \frac{\hat{\phi}_m^2}{24}
	\frac{\hat{\phi}_m^2+4\hat{\phi}_m(\hat{x}+\hat{y})+12\hat{x}\hat{y}}{\hat{\phi}_m+\hat{x}+\hat{y}}
\end{equation}

In case D (transient at exit) where the source is inside the dipole
while the observer is outside the dipole, the integrated CSR wake function is
given as:
\begin{eqnarray}
I_{csr} & = &\frac{\gamma r_c mc^2}{R}\Big( \frac{1}{\zeta}-\frac{2(\hat{\psi}+\hat{x}+\hat{\psi}^3/2+\hat{\psi}^2\hat{x})}
	{(\hat{\psi}+\hat{x})^2+(\hat{\psi}\hat{x}+\hat{\psi}^2/2)^2} \Big )
\end{eqnarray}
where $\hat{\psi}$, the normalized arc length from the source to the dipole exit,
can be found from the solution of the following equation:
\begin{equation}
\zeta = \frac{\hat{\psi}+\hat{x}}{2}+ \frac{\hat{\psi}^2}{24}\frac{\hat{\psi}+4\hat{x}\hat{\psi}}{\hat{\psi}+\hat{x}}
\end{equation}
The transient cases C and D at the exit of the bending dipole magnet are not included in
a drift element.

\begin{figure}[!htb]
   \centering
   \includegraphics*[angle=0,width=200pt]{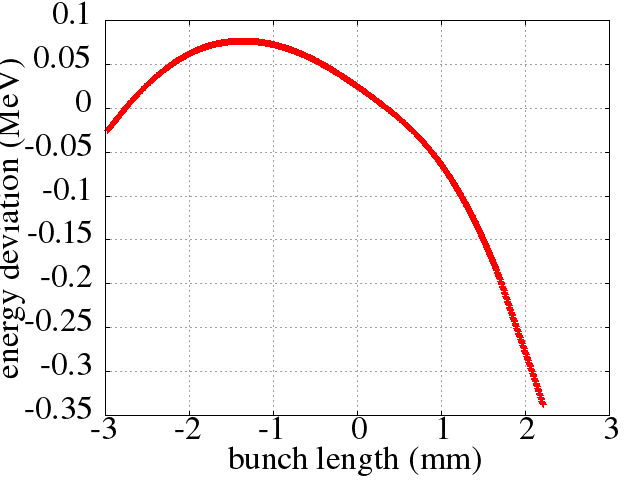}
   \includegraphics*[angle=0,width=200pt]{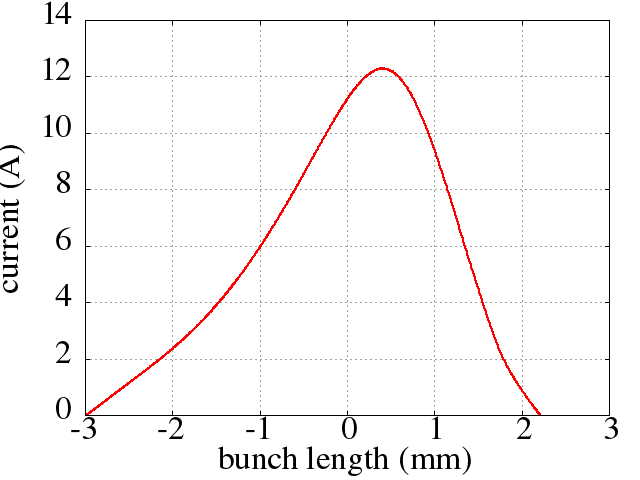}
\caption{Initial longitudinal phase space distribution (left) and
current profile (right) at the entrance of the linac.}
   \label{fig1}
\end{figure}

As a benchmark of the above model, we applied this model to the longitudinal beam dynamics
simulation of a LCLS-II linac design with 100pC charge and compared the 
longitudinal phase space and current profile from this one-dimensional model with those 
from the fully 3D multi-particle simulation using real number of electrons.
The LCLS-II is a high repetition rate (1 MHz) x-ray FEL
that will deliver photons of 
energy between 200 eV and 5 keV~\cite{paul,tor}. Figure~\ref{fig0} shows 
a schematic of LCLS-II superconducting RF linac. 
It  consists of  a laser heater (LH) to suppress microbunching instability, a section of superconducting linac L1 to accelerate the beam to about 270 MeV, 
a section of a third harmonic cavities (also called harmonic linearizer (HL) ) to 
linearize electron beam longitudinal phase space and decelerate the beam
down to about $230$ MeV, a bunch compressor BC1, a second section of 
superconducting linac L2 to accelerate the beam to about 1.6 GeV, another bunch compressor BC2, and a third section of superconducting linac L3 to accelerate the beam to about final 4 GeV, a long bypass transport line, and a 
magnetic kicker to spread the electron beam to a soft x-ray transport beam line and to a hard x-ray transport beam line. The superconducting linacs in all three sections are made of 1.3 GHz 9 cell superconducting cavities except the two cryomodules of 3.9 GHz third harmonic cavities right before the BC1 to linearize longitudinal phase space. 

\begin{figure}[!htb]
   \centering
   \includegraphics*[angle=0,width=200pt]{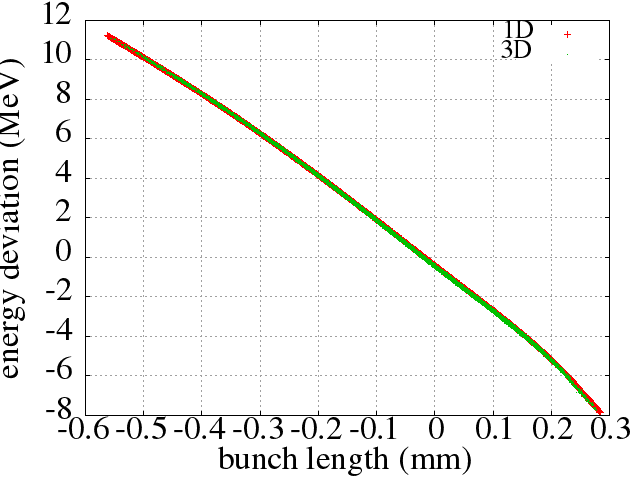}
   \includegraphics*[angle=0,width=200pt]{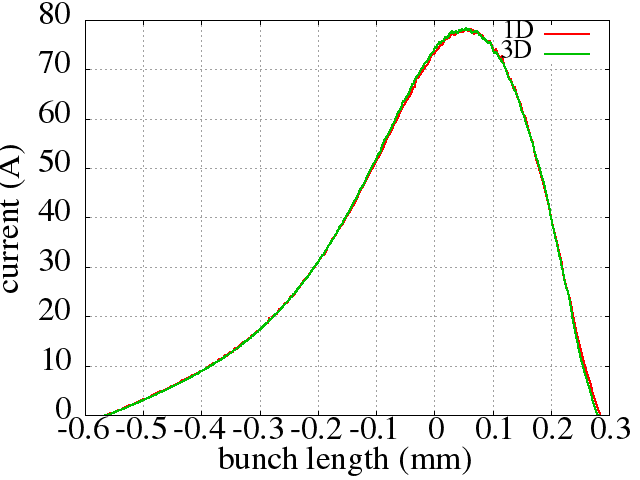}
\caption{Longitudinal phase space distribution (left) and
	current profile (right) after BC1 of the linac from the 1D model (red)
	and from the 3D model (green).}
   \label{fig2}
\end{figure}

\begin{figure}[!htb]
   \centering
   \includegraphics*[angle=0,width=200pt]{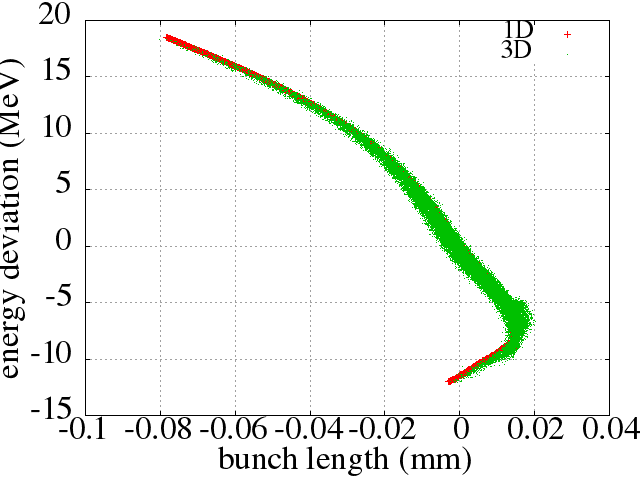}
   \includegraphics*[angle=0,width=200pt]{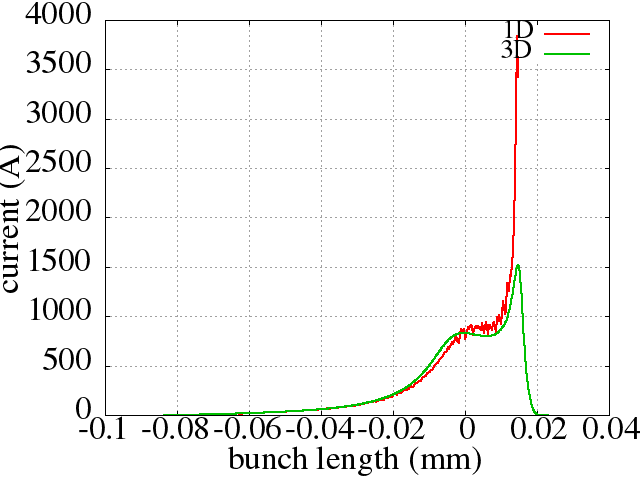}
\caption{Longitudinal phase space distribution (left) and
	current profile (right) after BC2 of the linac from the 1D model (red)
	and from the 3D model (green).}
   \label{fig3}
\end{figure}

\begin{figure}[!htb]
   \centering
   \includegraphics*[angle=0,width=200pt]{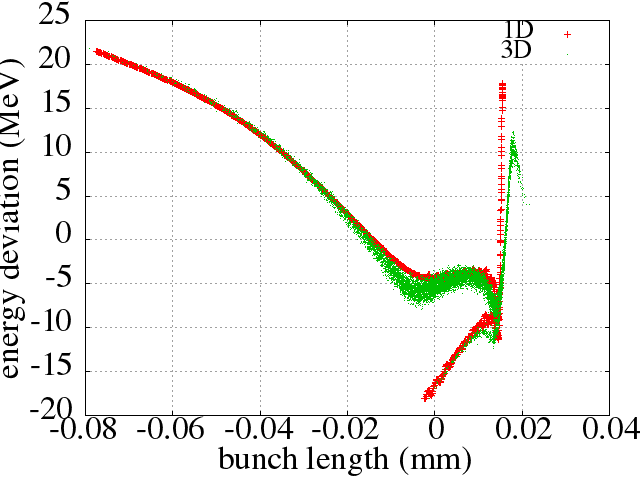}
   \includegraphics*[angle=0,width=200pt]{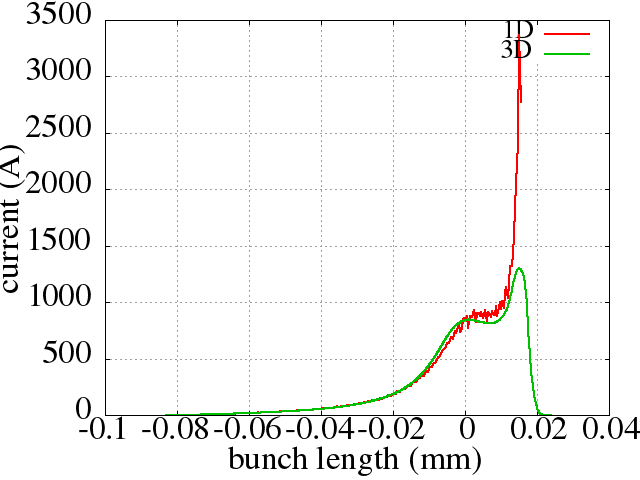}
\caption{Longitudinal phase space distribution (left) and
	current profile (right) at the entrance to undulator from the 1D model (red)
	and from the 3D model (green).}
   \label{fig4}
\end{figure}
Figure~\ref{fig1} shows the initial longitudinal phase space and current
profile at the entrance of the LCLS-II linac.
This initial beam distribution has an energy of about $100$ MeV coming out of 
a high repetition rate photoinjector.
A total $1024$ macroparticles (corresponding to $1024$ grid points/slices) 
that uniformly distributed along the bunch length coordinate are used in the 1D simulation.
The energy deviation of each macroparticle can be obtained from the
left longitudinal phase space plot of the Fig.~\ref{fig1}. The charge weight of each macroparticle
can be obtained from the right current plot, i.e. $w_i = I_i \delta z/c$, where
$I_i$ is the current at longitudinal bunch position $i$ and $\delta z$ is the 
longitudinal grid size.


Figure~\ref{fig2} shows the longitudinal phase space and current profile
after the magnetic bunch compression chicane BC1 from the tracking using
$1024$ macroparticles and the above one-dimensional (1D) longitudinal beam dynamics model
and the lumped elements and from the three-dimensional (3D) element-by-element 
tracking using the IMPACT code with the real number of electrons.
In the 1D tracking, the longitudinal space-charge effect
and the structure wakefields from the RF cavities are included
in the simulation. The CSR effect is applied only through the last bending
dipole magnet where the bunch length is short and the CSR effect is strong~\cite{chad}.
It is seen that both models agree with each other very well after BC1.

Figure~\ref{fig3} shows the longitudinal phase space and current profile
after magnetic bunch compression chicane BC2 from the
the above 1D longitudinal beam dynamics model
with lumped beam line elements and from the 3D element-by-element 
multi-particle tracking using the IMPACT code.
It is seen that the longitudinal phase space and the current profile from both
models agree with each other quite well. The large current spike
around the head of the beam from the 1D model is due to the fact that
the 1D model started with a distribution with zero uncorrelated energy
spread. Such uncorrelated energy spread in the 3D multi-particle simulation
smears the longitudinal phase space somewhat and reduces the 
current spike near the head of the beam. 
The longitudinal space-charge effect and the RF cavity structure wakefields
are included in the 1D tracking. The CSR effect was applied through 
the last bending magnet of the BC2.

After the bunch compressor BC2, the electron beam moves through 
another accelerating section (linac 3), a long transport
beam line, and a hard x-ray beam transport line,
before entering the undulator section. There are a number of dogleg
sections in the transport beam line. Here, we included the longitudinal
space-charge effect, the structure wakefield in linac3, the CSR effect
inside a long bending magnet with a large bending angle, and the
resistive wall wakefields. 
Figure~\ref{fig4} shows the final longitudinal phase space and current profile
at the end of accelerator beam delivery system, i.e. at the entrance
of undulator section from
the above 1D longitudinal beam dynamics model
with lumped elements and from the 3D element-by-element 
tracking using the IMPACT code.
The final longitudinal phase space and the current profile from the 1D model
agree with those from the 3D model quite well. The longitudinal 
phase space from the
1D model shows similar shape to that from the 3D model including the fold-over
particle distribution near the head of the electron beam. The current spike
around the head of the bunch from the 1D model 
is higher than the spike from the 3D simulation
due to the absence of the initial uncorrelated energy spread in the particle
distribution.



The benchmark between the above 1D longitudinal beam dynamics model using lumped
elements and the 3D element-by-element multi-particle simulation
shows good agreement between those two models. This gives us confidence in
applying this fast 1D model to final longitudinal phase space optimization
to attain high peak current and flat longitudinal phase space distribution.

\section{Multi-Objective Differential Evolution Optimization Algorithm}
In many accelerator applications, one needs to optimize more than one objective function. 
The problem of multi-objective optimization can be stated 
in the general mathematical form as: 
\begin{eqnarray}
 min \left \{ \begin{array}{l} 
	 f_1(\vec{x}) \\
                   \cdots \\
		   f_n(\vec{x})
                    \end{array}
                  \right.
	subject \ to \ g_i(\vec{x}) \leq 0, h_i(\vec{x}) = 0
\end{eqnarray}
Here, $f_1,\cdots,f_n$ are $n$ objective functions to be optimized,
$\vec{x}$ is a vector of control parameters, and $g_i$ and $h_i$ are constraints 
to the optimization.
The goal of multi-objective optimization is to find the Pareto front in the objective function solution space.
The Pareto optimal front is a collection of non-dominated solutions in the whole feasible solution space.
Any other solution in the feasible solution space will be dominated by those solutions on the Pareto optimal front. 
In the multi-objective optimization, a solution A is said to 
dominate a solution B if all components of A are at least as good as 
those of B (with at least one component strictly better). Here, a component
of A corresponds to one objective function value, i.e. $A_i = f_i(\vec{x})$.
The solution A is non-dominated if it is not dominated by any solution
within the group. 

Recently, we developed a new parallel multi-objective differential evolution 
algorithm with varying population size of each generation and with external storage to save
non-dominated solutions~\cite{qiangop3,qiangop4}. The use of variable population from
generation to generation is based on
the observation that during the early stage of evolution, 
the number of nondominated
solutions is small. There is no need to keep many dominated solutions in
the parent population. As the search evolves, more and more nondominated
solutions are obtained.
Those nondominated solutions are stored in an
external storage so that they can be used for selecting the new 
parent population. The advantage of using a variable population size
with external storage is to reduce the number of objective function evaluations
and to improve the speed of convergence.
The new algorithm is
summarized in the following steps: 
\begin{itemize}
\item Step 0: Define the minimum parent size, $NPmin$ and the maximum size, $NPmax$ of the parent 
            population. Define the maximum size of the external storage, $NPext$. 
\item Step 1: An initial $NPini$ population of parameter vectors are chosen quasi-randomly to 
           cover the entire solution space.
\item Step 2: Generate the offspring population using a unified
	differential evolution algorithm.
\item Step 3: Check the new population against the constraints.
\item Step 4: Combine the new population with the existing parent population from    
            the external storage. Non-dominated solutions ($Ndom$) are found from this group of solutions and       
            $min(Ndom, Next)$ of solutions are put back to the external storage. Pruning is used
            if $Ndom>Next$. $NP$ parent solutions are selected from this group of solutions for next generation 
            production. If $NPmin \leq Ndom\leq NPmax$, $NP = Ndom$. Otherwise, $NP=NPmin$ if 
            $Ndom<NPmin$ and $NP=NPmax$ if $Ndom > NPmax$. The elitism is emphasized
            through keeping the non-dominated solutions while the diversity is maintained 
            by penalizing the over-crowded solutions through pruning. 
\item Step 5: If the stopping condition is met, stop. Otherwise, return to Step 2.
\end{itemize}

The differential evolution algorithm is a simple but powerful method
for global parameter optimization~\cite{storn2,ali,price}. 
Compared with the other evolutionary
algorithms such as the genetic algorithm, the differential evolution 
algorithm makes use of the differences of parent solutions to
attain gradient information. This helps improve the convergence
speed of the algorithm in comparison to the genetic algorithm. 
Meanwhile, compared with the particle swarm method, the differential evolution
algorithm has a cross-over stage to enhance the diversity of solutions.
This helps the differential evolution algorithm to 
avoid converging to a premature solution.

%

The differential evolution algorithm generates new 
offsprings using two operations: mutation and crossover.
During the mutation stage, for each population member 
(target vector) $\vec{x}_{i}$, $i=1,2,3,\cdots, NP$
at generation $G$, a new mutant vector $\vec{v}_{i}$ is generated by following
a mutation strategy. 
A number of mutation strategies have been proposed for the conventional standard differential
evolution algorithm. 
The presence of multiple mutation strategies complicates the
use of the differential evolution algorithm. 
Recently, we proposed a single mutation expression that 
can unify most conventional mutation strategies used by
the differential evolution algorithm~\cite{qiangop2}.
This single unified mutation strategy can be written as:
\begin{eqnarray}
\vec{v}_{i} & = & \vec{x}_{i} + F_{1} (\vec{x}_{b} - \vec{x}_{i}) + F_{2} (\vec{x}_{r_1} - \vec{x}_{i}) + F_{3} (\vec{x}_{r_2}-\vec{x}_{r_3})
	\label{ude}
\end{eqnarray}
Here, the second term on the right-hand side of equation~(\ref{ude}) denotes
the contribution from the best solution found in the current generation,
the third term denotes the rotationally invariant contribution from the random
solution~\cite{price2},
and the fourth term is the same terms as that used in
the original differential evolution algorithm to account for
the contribution from the difference of parent solutions.
Those last two terms divert the mutated solution away from the best solution
and help to improve the algorithm's exploration of the decision parameter space.
The three parameters
$F_1$, $F_2$, and $F_3$ are the weights from each contribution.
This unified expression represents a combination of
exploitation (using the best found solution) and exploration (using
randomly chosen solutions) when generating the new mutant solution. 

Using the equation (\ref{ude}), the multiple mutation strategies of the standard differential evolution
algorithm can be included in a single expression. 
For example, the standard differential evolution algorithm such as DE/rand/1 can be 
attained by setting $F_1=0$, and $F_2=1$.
This new expression provides an opportunity to explore more broadly the space of mutation operators.
Using a different set of parameters
$F_1,F_2,F_3$, a new mutation strategy can be achieved. 
Moreover, by adjusting these parameters during the 
evolution,
the multiple mutation strategies and their combinations 
can be used during different stages of optimization. 
Thus, the unified mutation expression has the virtue of mathematical simplicity and also provides users with flexibility for broader exploration of different mutation strategies.

A crossover operation between the new generated mutant vector $\vec{v}_{i}$ and 
the target vector $\vec{x}_{i}$
is used to further increase the diversity of the new candidate solution.
This operation combines the two vectors into
a new trial vector 
$\vec{U}_{i}, i=1, 2, 3, \cdots, NP$, where the components of the trial vector are obtained from the components of $\vec{v}_i$ or $\vec{x}_i$ according to 
a crossover probability $Cr$.
In the binomial crossover scheme for a $D$ dimensional 
control parameter space, the new trial vector $\vec{U}_{i}$, $i=1, 2, \cdots, NP$ is generated using the following rule:
\begin{eqnarray}
\vec{U}_{i} & = & (u_{i1},u_{i2},\cdots,u_{iD}) \\
u_{ij} & = & \left \{ \begin{array}{ll} 
                   v_{ij}, & {\rm if} \ \ {\rm rand}_j \le Cr \ \ {\rm or} \ j={\rm mbr}_i \\
                   x_{ij}, & {\rm otherwise}
                    \end{array}
                  \right.
\end{eqnarray}
where rand$_j$ is a randomly chosen real number in the interval $[0, 1]$, and 
the index $mbr_i$ is a randomly chosen integer in the range $[1, D]$.  This ensures 
that the new trial vector contains at least one component from the 
new mutant vector. 

During the mutation stage and the crossover stage at generation, each individual 
solution $\vec{x}_{i}$, $i=1,2,3,\cdots, NP$ has a set of control parameters $F_{1,i},F_{2,i},F_{3,i}$ and $Cr_i$ associated with it.  
Before generating a new mutant solution using the unified differential
evolution expression~(\ref{ude}), a set of control parameters 
$F_{1,i},F_{2,i},F_{3,i}$ and $Cr_i$
are calculated as:
\begin{eqnarray}
F_{j,i} & = & F_{jmin}+r_{ji} (F_{jmax}-F_{jmin})  \\
Cr_i	& = & Cr_{min}+r_{i} (Cr_{max}-Cr_{min})
\end{eqnarray}
where $r_{ji},r_{i}, j=1,2,3$  are uniform random values 
in the interval $[0,1]$, $F_{jmin}$ and $F_{jmax}$ for $j=1,2,3$ are the minimum
and the maximum allowed values of those control parameters,
$Cr_{min}$ and $Cr_{max}$ are the minimum and the maximum cross-over probability.
The values of $F_{jmin}$ and $F_{jmax}$ are set to $0$ and $1$ respectively in this
study. We also set $Cr_{min}=0.5$ and $Cr_{max}=1$.
The selection of these values is based on the consideration that the 
various conventional differential evolution mutation strategies
can be covered by the settings of those parameters, and in the literature, $F_3$ 
is rarely greater than one. 

\section{Longitudinal beam dynamics optimization of a LCLS-II design}

We applied the above fast one-dimensional longitudinal beam dynamics model together 
with the multi-objective
optimization algorithm to an existing LCLS-II design optimization.
The two objective functions in this optimization are the final negative fraction of charge and
rms energy spread inside a longitudinal phase space window. Here, the window's bunch length is set between
-$5$ microns and $5$ microns, while the energy spread is set between -$8$ MeV
and $8$ MeV. The smaller the negative fraction of charge, the higher
the peak current will be. The smaller the rms energy spread inside the window, the flatter the longitudinal
phase space will be. In general, the higher peak current might result in the larger energy spread.
By simultaneously optimizing those two conflict objectives, we hope to attain the best
achievable solutions, i.e. the Pareto optimal front. 
The high peak current and flat longitudinal phase space improves the x-ray FEL
radiation power and reduces the radiation bandwidth. 

\begin{figure}[!htb]
   \centering
   \includegraphics*[angle=0,width=200pt]{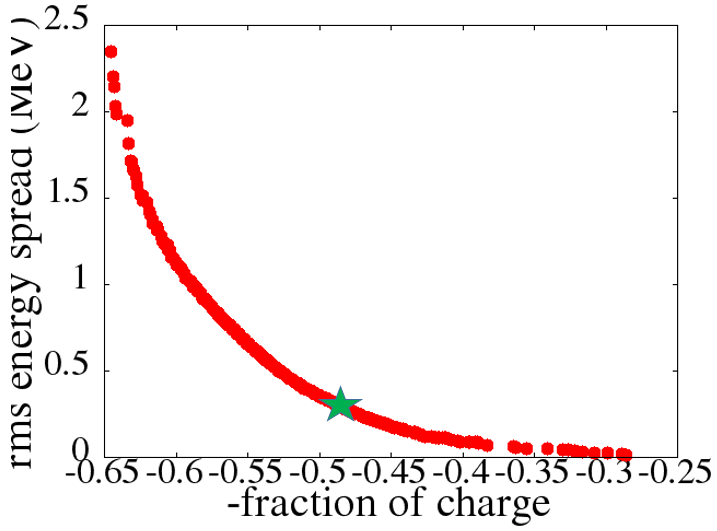}
   \caption{The Pareto front of longitudinal phase space optimization.}
   \label{fig5}
\end{figure}

We used $10$ control parameters
of the linear accelerator in this optimization. These $10$ control parameters 
in the LCLS-II linac are the linac
section one RF cavity accelerating gradient amplitude and RF phase, 3rd harmonic cavity amplitude and phase, bending angle in bunch compressor one, linac section two RF cavity amplitude and phase, bending
angle in bunch compressor two, and linac three RF cavity amplitude and phase.
The maximum RF cavity accelerating gradient is constrained around $16MV/m$ and the final energy
around $4$ GeV.
The initial population size is $256$. The minimum and the maximum population
size are set as $128$ and $1024$ respectively. The maximum number of 
nondominated solutions in the external storage is set as $2000$.
The Pareto front converges after about $460$ generations with about $76$ 
thousand objective function evaluations. 
It took one and half hour computing time on $64$ Intel Xeon Phi Processor $7250$ at
National Energy Research Supercomputing Center.
A major section of the Pareto front (with fraction of charge less than $0.6$)
has converged after one hour of computing time after $360$ generations
with $51$ thousand function evaluations.
Figure~\ref{fig5} shows the Pareto optimal front of the final rms energy spread and
the negative fraction of charge inside the defined window from the 
above multi-objective longitudinal beam dynamics optimization.
This optimal front suggests that the more charge inside the core of the beam, 
the larger correlated energy spread the longitudinal phase space will be.

\begin{figure}[!htb]
   \centering
   \includegraphics*[angle=0,width=200pt]{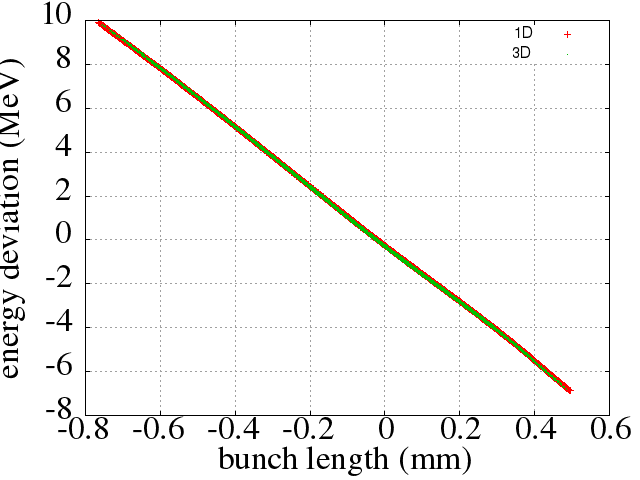}
   \includegraphics*[angle=0,width=200pt]{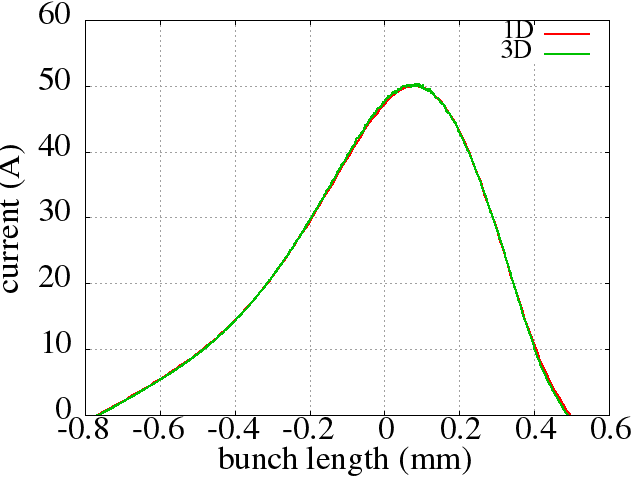}
\caption{Longitudinal phase space distribution (left) and
	current profile (right) after BC1 of the LCLS-II linac from the 1D model (red)
	and from the 3D model (green) using an optimal solution settings.}
   \label{fig6}
\end{figure}

\begin{figure}[!htb]
   \centering
   \includegraphics*[angle=0,width=200pt]{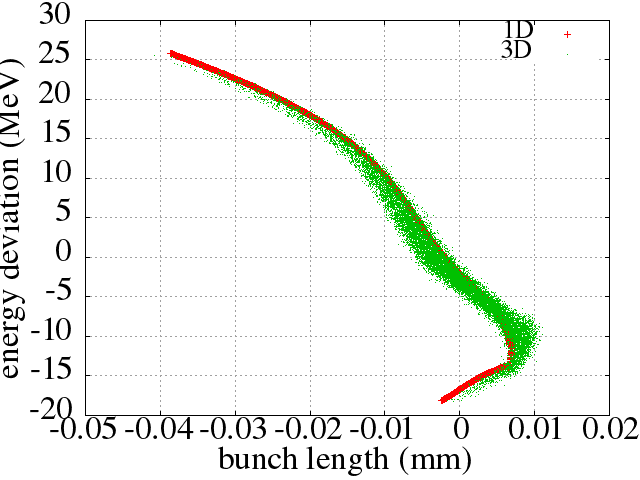}
   \includegraphics*[angle=0,width=200pt]{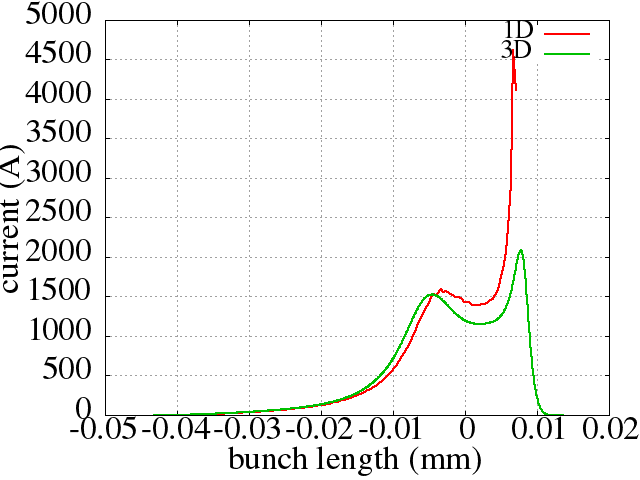}
\caption{Longitudinal phase space distribution (left) and
	current profile (right) after BC2 of the linac from the 1D model (red)
	and from the 3D model (green) using an optimal solution settings.}
   \label{fig7}
\end{figure}

\begin{figure}[!htb]
   \centering
   \includegraphics*[angle=0,width=200pt]{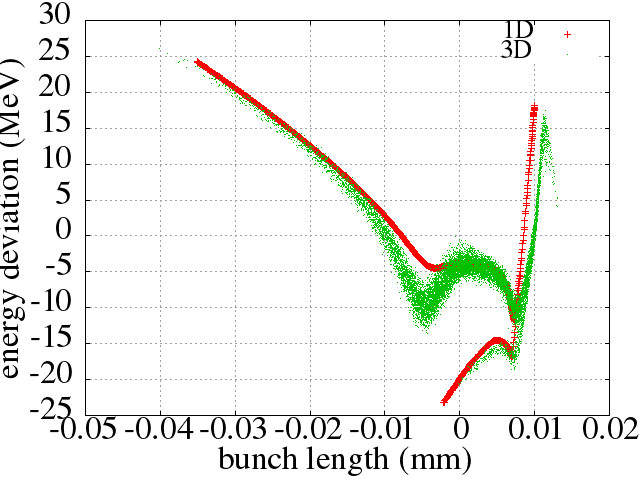}
   \includegraphics*[angle=0,width=200pt]{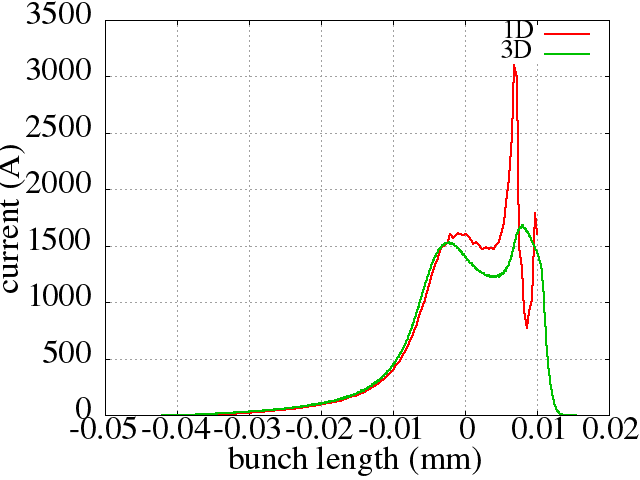}
\caption{Longitudinal phase space distribution (left) and
	current profile (right) at the entrance to undulator from the 1D model (red)
	and from the 3D model (green) using an optimal solution settings.}
   \label{fig8}
\end{figure}

From those optimal solutions, we selected one optimal solution, the green star
in Fig.~\ref{fig5}. For this solution, the optimized linac one RF cavity accelerating gradient amplitude and phase
are $11.5$ M/m and $-14.9$ degrees respectively, harmonic linearizer RF cavity
amplitude and phase are $9.0$ M/m and $153.6$ degrees respectively, bunch compression one
bending angle is $0.10$ radian, linac two RF cavity amplitude and phase are
$16.1$ M/m and -$30.4$ degrees, bunch compressor two bending angle is $0.047$ radian,
linac three amplitude and phase are $16.0$ M/m and $0.0$ degrees.
With those parameter settings of the LCLS-II linac, we ran the fast 1D longitudinal 
beam dynamics simulation and the fully 3D element-by-element
multi-particle simulation using the IMPACT code. The longitudinal phase spaces
and current profiles after BC1, after BC2, and at the entrance to undulator are
given in Figs.~\ref{fig6}-\ref{fig8}. It is seen that the predictions of the 1D
longitudinal beam dynamics model agree with those of the 3D model quite well.
There is about a factor of four compression after the first bunch compression chicane.
After the second bunch compressor chicane, the electron beam is further compressed by more
than a factor of $20$.
The optimized solution results in a final core peak current greater than $1.2$ kA,
which is about $50\%$ improvement from the previous design of around $800$ A
peak current.
Such a higher core peak current can result in higher x-ray FEL radiation power.

%

\section{Conclusions}

In this paper, we proposed a lumped one-dimensional longitudinal beam dynamics model that includes
nonlinear effects from drift, RF acceleration, magnetic bunch compression, longitudinal
space-charge effect, CSR effect, and structure and resistive wall wakefields, for fast
longitudinal phase space optimization in x-ray FEL linear accelerators. 
Benchmarking using a LCLS-II design, this simplified model shows good agreement with the
fully 3D element-by-element multi-particle tracking using the IMPACT code.
This model, implemented in a recently developed
multi-objective differential evolution optimization program, 
provides a fast computational tool for the longitudinal phase space optimization
and resulted in an improved solution on the existing LCLS-II design.
In the future study, we plan to apply this tool to the optimization of 
using control parameters from 
both the linac parameter setting
and the initial longitudinal phase space.
Those optimization results will help set up the requirements for the electron beam longitudinal
distribution out of a photoinjector in order to achieve the optimal final electron beam
quality at the entrance of undulator. 



\section*{ACKNOWLEDGEMENTS}
We would like to thank the LCLS-II physics design team for the LCLS-II application
study.
This work was supported by the U.S. Department of Energy under Contract No. DE-AC02-05CH11231
and used computer resources at the National Energy Research
Scientific Computing Center.

\end{document}